\documentclass[aps,showpacs,superscriptaddress,groupedaddress]{revtex4-1}  
\usepackage{graphicx}  % needed for figures
\usepackage{dcolumn}   % needed for some tables
\usepackage{bm}        % for math
\usepackage{amssymb}   % for math
\usepackage{xcolor}
\usepackage{graphics} 
\usepackage{hyperref}

\usepackage{epsfig} 
\input{epsf}

\def\beq{\begin{equation}}
\def\eeq{\end{equation}}
\def\be{\begin{equation}}
\def\ee{\end{equation}}
\def\bea{\begin{eqnarray}}
\def\eea{\end{eqnarray}}

\newcommand{\MeV}{{\rm MeV}}				% MeV
\begin{document}

\title{ Equation of state for strange quark matter: Linking the Nambu--Jona-Lasinio model to  perturbative QCD }
 
\author{Marcus Benghi Pinto} \email{marcus.benghi@ufsc.br}
\affiliation{Departamento de F\'{\i}sica, Universidade Federal de Santa
  Catarina, Florian\'{o}polis, SC 88040-900, Brazil}

 \begin{abstract} 
Neutron star constraints  and {\it ab initio} pQCD evaluations  require the EoS representing cold quark matter to be stiff at intermediate baryonic densities and soft at high-$n_B$. Here, I suggest that the three flavor NJL model with a  density dependent repulsive coupling, $G_V(\mu)$,  can generate an EoS which interpolates between these two   regimes.  Such an interpolation requires repulsion to start decreasing with the chemical potential just after chiral transition takes place. The conjecture behind this mechanism is that repulsion should be necessary only as long as the quark condensates, which dress the effective masses, have non-vanishing values. This assumption  guarantees that an initially hard EoS suffers a conspicuous change of slope at ${\cal E} \simeq 0.7 \,{\rm GeV fm^{-3}}$ converging to the pQCD results at higher energy densities. Then, the speed of sound naturally reaches a non-conformal maximum at $n_B = 3.23 \, n_0 = 0.52 \, {\rm fm}^{-3}$ while  the trace anomaly remains positive for all densities, in agreement with recent investigations.
These  non-trivial results {\it cannot} be   simultaneously obtained when $G_V$ vanishes or has a fixed value.  Therefore, the  simple model proposed here is able to    link the (non-perturbative) region of intermediate densities  to the region where pQCD becomes reliable. 
\end{abstract}
\maketitle 
 \section{Introduction}
 
 Effective quark models, such as the Nambu--Jona-Lasinio model (NJL) \cite {njl} and the MIT bag model \cite{mit1,mit2,mit3} capture some of the most representative characteristics of quantum chromodynamics (QCD), like confinement and chiral symmetry respectively \cite{buballa}. As a consequence, they are widely used to describe the thermodynamics of strongly interacting matter in regions of the phase diagram which are currently unaccessible to {\it ab initio} evaluations. Nowadays, the corner of low densities and high temperatures can be well described by first principle evaluations based on lattice QCD simulations (LQCD). However, due to the well documented sign problem, LQCD is not yet in position to describe the corner of low temperatures and finite baryonic densities which concerns neutron stars (NSs).  
 In this case the QCD equation of state (EoS)  describing cold and dense strongly interacting matter can be reliably evaluated only in regimes where the baryonic density ($n_B$) is  very low or extremely high. In the limit of low densities, chiral effective theory (CET) \cite {cet1,cet2}provides an accurate EoS up to about $n_B \lesssim \; 2 \, n_0 \equiv n_{\rm CET}$ ($n_0=0.16\,{\rm fm}^{-3}$) so that the region composed by  hadronic matter may be well described. At the other extremum, perturbative QCD (pQCD) \cite{pqcd1,pqcd2,pqcd3} gives a reliable equation for $n_B \gtrsim 40 \,n_0 \equiv n_{\rm pQCD}$, when quarks and gluons represent the relevant degrees of freedom \cite{kojo}. However,  at the  intermediate  range  $ 2n_0 \lesssim n_B \lesssim 8 n_0$, which concerns NSs, $\alpha_s$ is still high so that non-perturbative techniques and/or model approximations are generally employed.  Within this region the presence of quark matter in massive NSs was recently found \cite {nature} to be linked to the behavior of the speed of sound, $V_s$. The investigation performed in Ref. \cite {nature}  suggests that if the conformal bound $V_s^2 \le 1/3$ is not strongly violated massive neutron stars should  have sizable quark-matter cores. Moreover, the recent discovery  of NSs whose estimated masses are about twice the value of the solar mass \cite{astro1, astro2,astro3}   and the theoretical predictions on the maximum (gravitational) mass performed in Refs.  \cite{measure1, measure2, measure3,measure4,measure5} favor  an stiff EoS with $V_s^2 >1/3$ at $n_B > n_0$. In this case, recent simulations \cite {sinansimulation, michal, weisesimulation} indicate that  $V_s^2$ is a non-monotonic
function of $n_B$ , which in turn suggests the existence of at least one local maximum where $V_s^2 > 1/3$. Together, all of these findings  constrain the EoS to be initially stiff (so that $V_s^2 >1/3$)  before softening, at intermediate densities, to finally meet the pQCD predictions at high-$n_B$.  As it is well known, when effective quark models are being employed  the inclusion of a repulsive vector channel, parametrized by $G_V$,  generates a harder EoS in most cases \cite{fukushima1,fukushima2}. However, a drawback is that such an equation remains stiff at higher densities so that the conformal limit, observed by pQCD, cannot be attained. On the other hand, when $G_V=0$,   asymptotic convergence to pQCD is observed but the EoS is far too soft to cope with NSs constraints at lower densities. 

One way to circumvent this problem is to  assume that  $G_V$ is density dependent  as recently proposed in Ref. \cite {letter}, where the two flavor NJL model has been considered. There, it  has been suggested that the repulsion among (dressed) quarks is important only up to the point where the chiral transition occurs so that repulsion among (bare) quarks should be negligible. In Ref. \cite {letter}, the running of $G_V$ was modelled by a simple ansatz which interpolates between a regime where repulsion is high (the EoS is stiff) and a regime where repulsion is low (the EoS is soft). Thanks to this property the  two-flavor NJL model with a running $G_V(\mu)$ predicted \cite{letter} a non-monotonic behavior for $V^2_s$  implying that the  existence of a peak,  at $n_B \simeq 3.25 n_0$, can be conciliated with  pQCD  predictions at asymptotically high baryonic densities. Physically, these results indicate that repulsion should be necessary only as long as the quark condensates (which are directly related to the NJL quark self energies) exist. 

Since strangeness may play an important role when describing more realistic situations the present work contemplates an extension to the case where this degree of freedom is present. With this purpose, the three flavor NJL model with a repulsive channel will be considered here as a prototype to describe cold strange quark matter. As it will be shown, also in this case a density dependent $G_V$ allows us for the presence  of a non-conformal bump in $V_s^2$ at $n_B = 3.23 \, n_0 = 0.52 \, {\rm fm}^{-3}$ (in agreement with Ref. \cite {michal}) while the trace anomaly remains  positive. This  rather non-trivial result  supports a recent claim \cite{fukushimatrace} which states that the presence of a non-conformal peak in $V_s^2$  is not necessarily in tension with the trace anomaly being positive for all densities. 
Concerning the EoS another important result obtained here predicts a prominent change of slope taking place at ${\cal E} \simeq 0.7\, {\rm GeV fm^{-3}}$, in agreement with what is  observed in  Refs. \cite {nature, michal}. These predictions indicate that the modified three flavor NJL model discussed in this investigation may contribute to describe the QCD EoS at intermediate baryonic densities. 

The paper is organized as follows. In the next section  the basic results for the three flavor NJL model are reviewed.
The possible density dependence of the repulsive vector interaction is presented in Sec.~\ref{sec3}.   Numerical results associated with
the relevant thermodynamical quantities are generated and discussed in
Sec.~\ref{sec4}. The conclusions are  presented in
Sec.~\ref{sec5}.

\section{The $N_f=2+1$ NJL model: standard results}
\label{NJL3}

In the presence of a repulsive vector channel  the standard three-flavor version of the NJL model can be written as \cite{fukushima1,fukushima2,campoB}

\begin{equation}
\mathcal{L}\ =\ {\bar\psi}(i\gamma_\mu\partial^\mu-m)\psi
+G_S\sum_{a=0}^8 \left[({\bar\psi}\lambda^a\psi)^2
	+(\bar{\psi}i\gamma_5\lambda^a\psi)^2\right] 
-K \left\{{\rm det}_f[\bar\psi(1+\gamma_5)\psi] 
	 +{\rm det}_f[\bar\psi(1-\gamma_5)\psi]\right\} - G_V ({\bar\psi}\gamma^\mu \psi)^2\ ,
\label{njl3}
\end{equation}
where $\psi=(u,d,s)^T$ denotes a quark field with three flavors 
(and three colors), 
and $m= {\rm diag}_f(m_u,m_d,m_s)$ is the corresponding mass matrix. 
Setting $m_u=m_d \equiv m \ne m_s$ implies  that isospin symmetry is observed 
while the SU(3) flavor symmetry is explicitly broken. 
The eight Gell-Mann matrices are represented by $\lambda^a$ ($a=1,...,8$) 
and $\lambda^0=\sqrt{2/3}\,\boldsymbol{I}$. 
In $3+1\, d$ the NJL model is composed by irrelevant operators so that the couplings $G_S$, $G_V$  and $K$ respectively have canonical dimensions [-2], [-2] and [-5] implying that the model is non-renormalizable. Here, the (ultra violet) divergent integrals  will be regularized by a sharp non-covariant cut-off, $\Lambda$, whose numerical value is set by phenomenological inputs. 
For the numerical analysis I adopt the parameter values of Ref.\cite{rkh}
which are $m=5.5\,\MeV$, $m_s=140.7\,\MeV$, $G\Lambda^2=1.835$, 
$K \Lambda^5=12.36$, and $\Lambda= 602.3\,\MeV$. 
Then, at $T=0$ and $\mu_f=0$, one reproduces $f_\pi=92.4\,\MeV$, 
$m_\pi=135\,\MeV$, $m_K=497.7\,\MeV$, and $m_{\eta^\prime}=960.8\,\MeV$. 
For the quark condensates one then obtains $\sigma_u=\sigma_d=-(241.9\,\MeV)^3$, 
and $\sigma_s=- (257.7\,\MeV)^3$. Fixing $G_V$ poses and
additional problem since this quantity should be determined by considering the $\rho$ meson mass which, in general, happens to be
higher than the maximum energy scale set by $\Lambda$. In this situation, most authors adopt values between $0.25 G_S$ and  $0.5 G_S$ (see Ref. \cite {tulio} for more details). Here, the value $G_V=G_S/3$ will be adopted when dealing with a {\it fixed} vector coupling \cite {tulio,sugano}.
Note that to assure rotational invariance only the zeroth component of the vector channel contributes so that, at the mean field level, the chemical potential gets shifted as \cite{buballa, fukushima1,fukushima2}
\begin{equation}
{\tilde \mu}_f = \mu_f  - 2 G_V \sum_f n_f \,,
\label{mutilde}
\end{equation}
with  $n_f$ representing the quark number density {\it per flavor} \cite{buballa, fukushima1,fukushima2,tulio}. 
At $T=0$, a standard mean field approximation (MFA) evaluation yields the following  result \cite{buballa,tulio}
\begin{equation}
n_f = \frac{N_c}{3\pi^2} p_{F,f}^3 \;,
\end{equation}
where the effective Fermi momentum is just $p_{F,f} = \sqrt { {\tilde \mu}_f^2 - M_f^2}$.  The quark effective masses are given by  \cite{buballa}
\begin{equation}
M_f = m_f - 4 G_S \sigma_f + 2 K \sigma_j \sigma_k\,,
\label{mass}
\end{equation}
where $\sigma_f = \langle {\overline \psi} \psi \rangle_f$ represents the quark condensate for a given flavor
\begin{equation}
\sigma_f = - \frac{N_c}{2\pi^2} M_f \left [ \Lambda p_{\Lambda,f} - M_f^2 \ln \left ( \frac{\Lambda + p_{\Lambda,f}}{M_f} \right)  \right ] 
+ \frac{N_c}{2\pi^2} M_f  \left [ {\tilde \mu}_f p_{F,f} - M_f^2 \ln \left ( \frac{{\tilde \mu}_f + p_{F,f}}{M_f} \right)  \right ] \,,
\end{equation}
with $p_{\Lambda,f} = \sqrt { \Lambda^2 + M_f^2}$. The effective Fermi momentum, $p_{F,f}$, is then determined by solving Eqs. (\ref{mutilde}) and (\ref{mass}) simultaneously.

Having the quark number density, $n=\sum_f n_f $, one  can express  the squared speed of sound in terms of  the baryonic number susceptibility, $\chi_B = d n_B/d\mu_B$, as 
\begin{equation}
V_s^2 = \frac{n_B}{\mu_B \chi_B} \,,
\label{sound}
\end{equation}
where   $\mu_B=\sum_f \mu_f$  and $n_B = n/3$. For simplicity, the present application  concerns the case of symmetric strange quark matter only so that one can now set $\mu_u = \mu_d =\mu_s \equiv \mu$. In the present context this choice can be further justified by recalling that three-flavor symmetric matter, at the high density limit, meets the conditions of $\beta$-stability and charge neutrality which are usually required to describe NSs \cite{constanca}. 

Then, at finite chemical potential and zero temperature, the pressure versus chemical potential relation for quark matter can be obtained from \cite{fukushima2, zong}
\begin{equation}
P(\mu) = P(0) + \int_0^\mu n(\nu) d\nu \,,
\label{pressao}
\end{equation}
 where $P(0) = $ is the vacuum pressure.  Notice that within the present approach the vector coupling, $G_V$, turns out to be density dependent so that one needs to be careful in order to preserve thermodynamic consistency. In general, there are two ways in which one can proceed. One possibility is to first evaluate the pressure and then  redefine this quantity so that the number density can be consistently obtained by deriving  the pressure with respect to the chemical potential. In this case  extra terms, such as $(\partial G_V/\partial \mu)(\partial P/\partial G_V)$,  get compensated thanks to the redefined pressure.  Within the second possibility, which is the one I adopt here, the quark number density is evaluated before being  numerically integrated  so that the $\mu$ dependence of $G_V$ is automatically accounted for (see Ref. \cite {newtulio} for more details). Then, from  Eq. (\ref{pressao}), one can determine  the energy density, ${\cal E} = - P + \mu_B n_B$, the trace anomaly, $\Delta = {\cal E} - 3P$, as well as the conformal measure, ${\cal C} = \Delta/{\cal E}$.

\begin{figure}
\centerline{ \epsfig{file=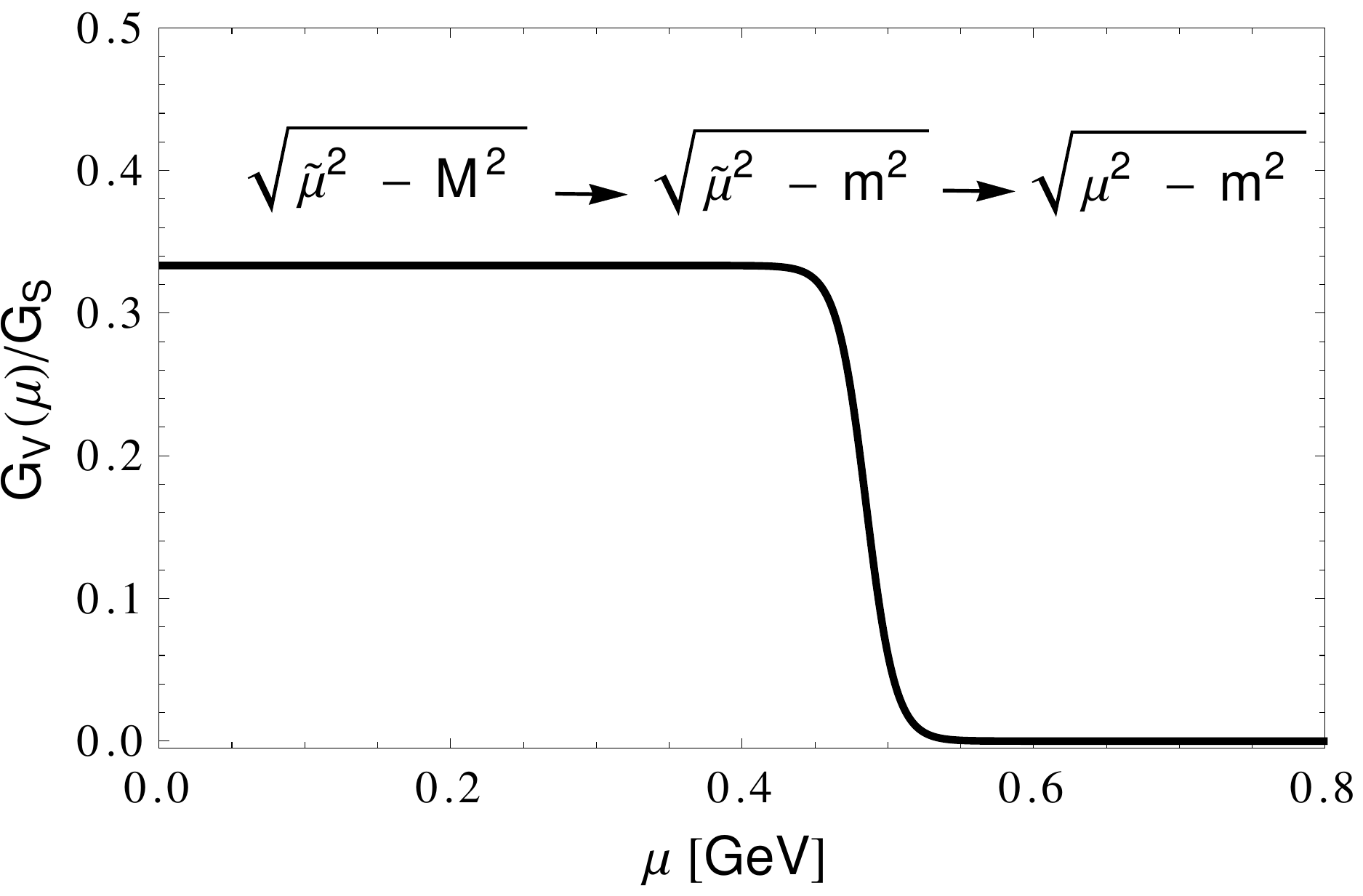,width=0.45\linewidth,angle=0}}
\caption{ Running repulsive coupling $G_V(\mu)$, in units of $G_S$, as a function of the quark chemical potential, $\mu$. The Fermi momentum corresponding  to each distinct region  is shown for reference. }
\label{Fig1}
\end{figure}

\begin{figure}
\centerline{ \epsfig{file=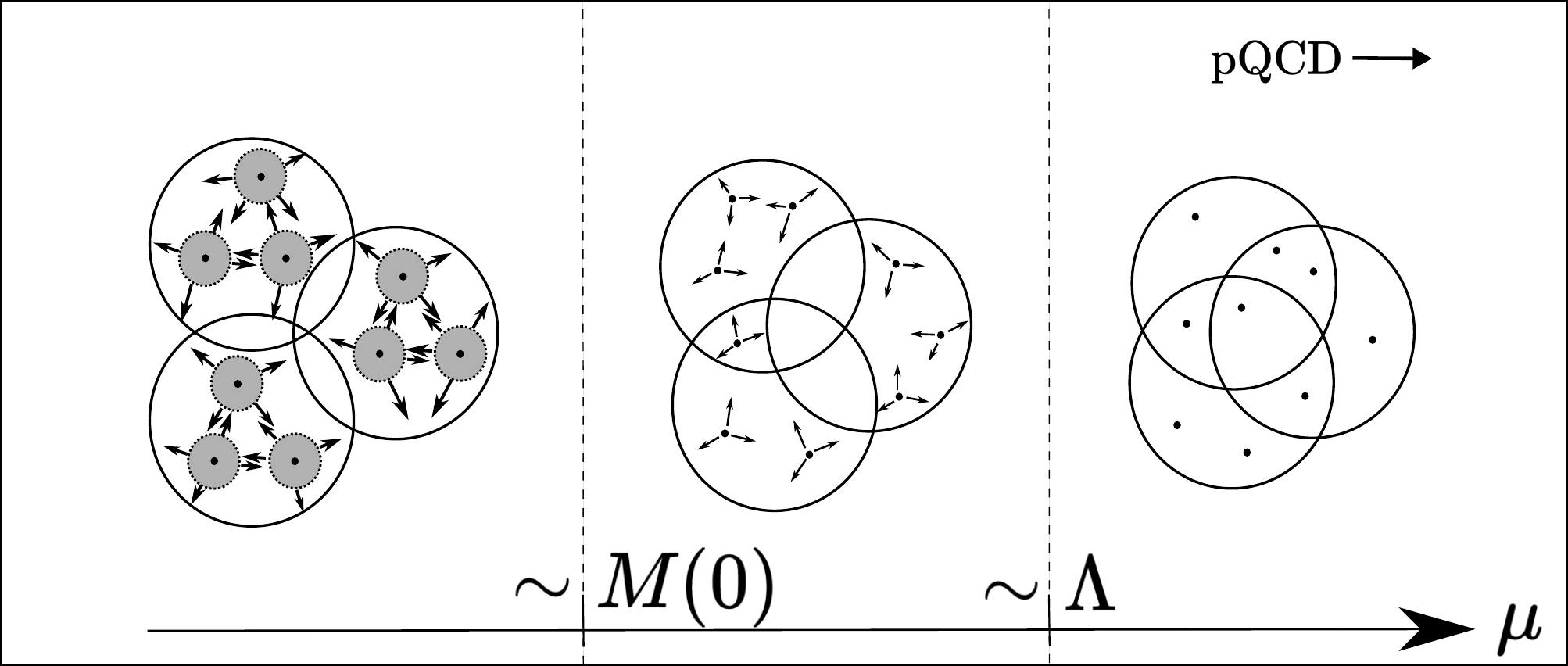,width=0.65\linewidth,angle=0}}
\caption{A pictorial representation of the physical conjecture driving the running of $G_V(\mu)$. Quarks with large effective masses (represented by the gray area) tend to strongly repel each other as compression increases. After the chiral transition occurs (at $\mu \sim M(0)$), and the masses tend to their bare values, quarks can be further compressed without repelling each other. Therefore, repulsion should be necessary only as long as the quark condensates (which dress the NJL masses, Eq. (\ref{mass})) exist ($\mu \lesssim \Lambda$). }
\label{Fig2}
\end{figure}

\begin{figure}
\centerline{ \epsfig{file=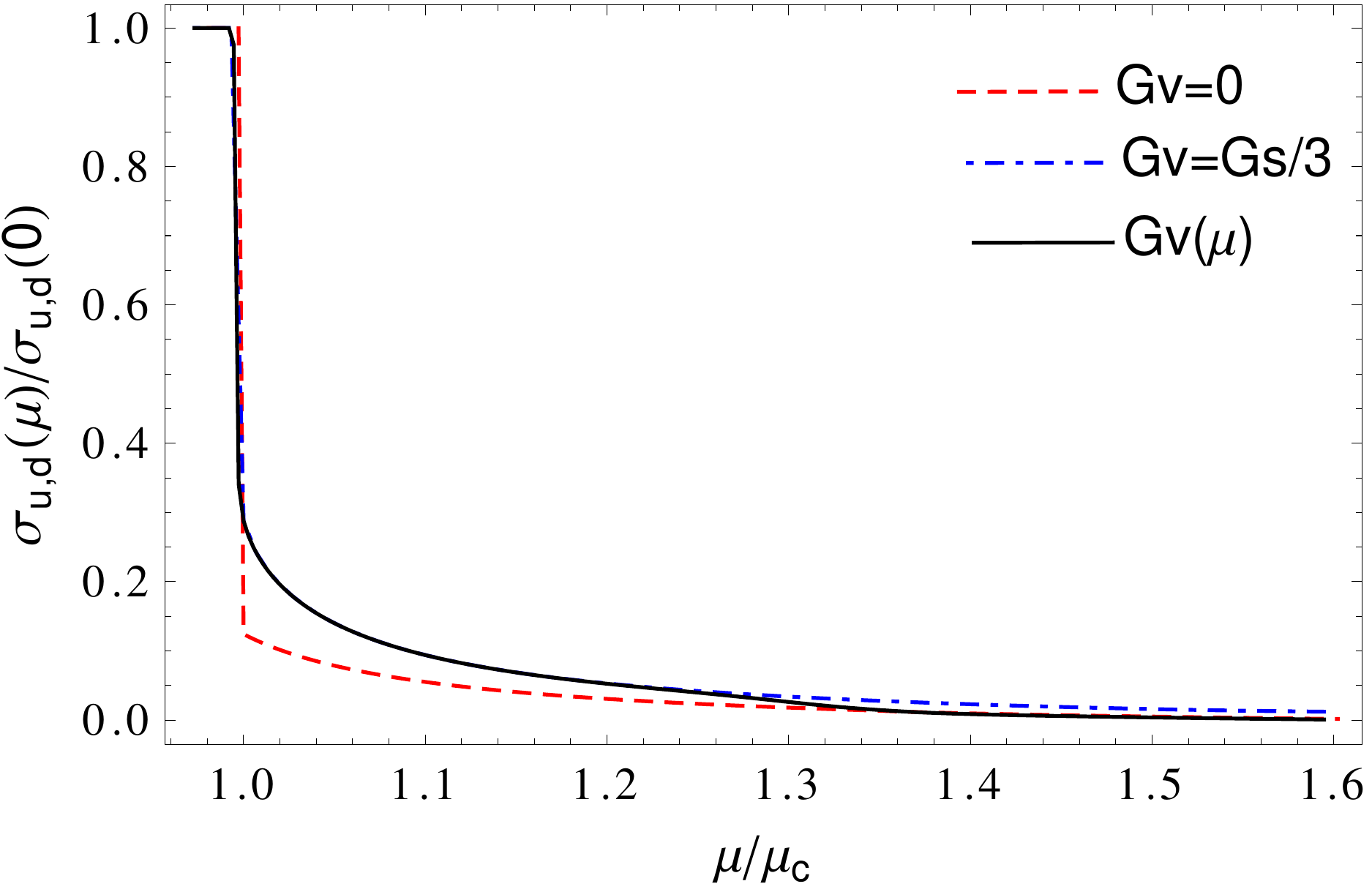,width=0.45\linewidth,angle=0}}
\caption{(Light flavor) quark condensate, $\sigma_{u,d}(\mu)$ normalized by $\sigma_{u,d}(0)$, as a function of $\mu/\mu_c$. For the cases $G_V=G_S/3$ and $G_V(\mu)$  the coexistence quark chemical potential is  $\mu_c = 0.368\, {\rm GeV}$ while for $G_V=0$ it reads $\mu_c=0.361 \, {\rm GeV}$. Up to  about $\mu = 1.3 \, \mu_c$ the $G_V(\mu)$ prediction agrees with the $G_V=G_S/3$ curve. After that it converges towards the $G_V=0$ curve, as expected.}
\label{Fig3}
\end{figure}

\begin{figure}
\centerline{ \epsfig{file=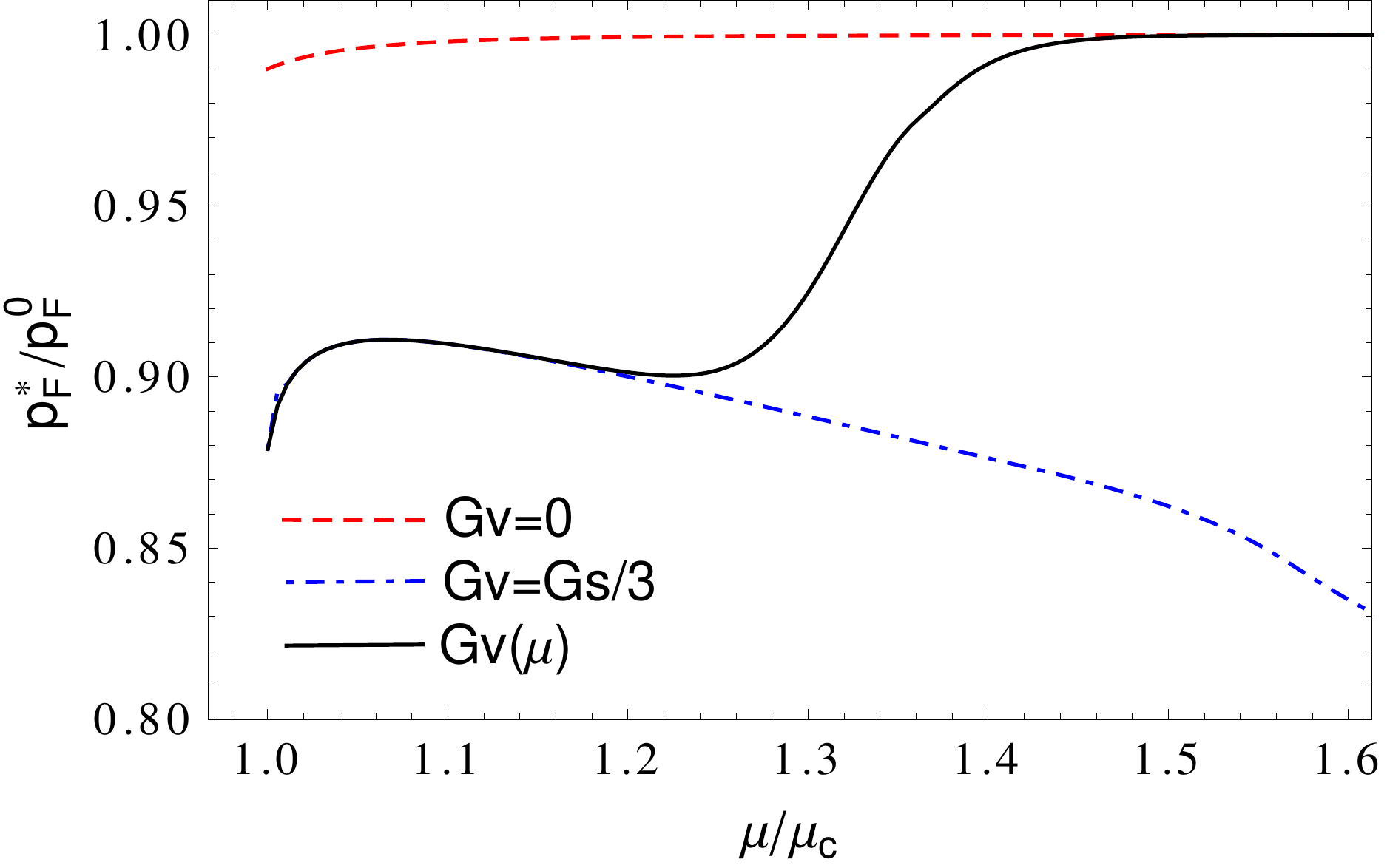,width=0.45\linewidth,angle=0}}
\caption{Effective Fermi momentum for light quarks, $p_F^*=\sqrt{{\tilde \mu}^2 - M^2}$, normalized by  $p_F^0 = \sqrt{\mu^2 - m^2}$, as a function of $\mu/\mu_c$. For the cases $G_V=G_S/3$ and $G_V(\mu)$ the coexistence quark chemical potential is  $\mu_c = 0.368\, {\rm GeV}$ while for $G_V=0$ it reads $\mu_c=0.361 \, {\rm GeV}$. The $G_V(\mu)$ curve interpolates between those predicted by the cases $G_V=G_S/3$ and $G_V=0$.}
\label{Fig4}
\end{figure}

\section{Interpolating between softness and stiffness}
\label{sec3}

Let me now discuss how to tune $G_V(\mu)$ so as to  obtain an EoS which  interpolates between the stiff and the soft regimes. When dealing with symmetric quark matter one can further simplify the notation by  setting   $M_u=M_d\equiv M$ as previously done for $m_u$ and $m_d$. Using these definitions and taking $G_V=0$  one can write the Fermi momentum for a light flavor, in symmetric matter, as $ \sqrt{\mu^2 - M^2}$. As compression increases the quark condensates decrease and the chiral transition sets in ($M \to m$) so that the Fermi momentum changes as $\sqrt{\mu^2 - M^2} \to p_F^{0} = \sqrt{\mu^2 - m^2}$, where $p_F^{0}$ represents the case of free (bare) quarks considered within pQCD. Now, when  $G_V$  is fixed chiral symmetry (partial) restoration  implies that $\sqrt{{\tilde \mu}^2 - M^2} \to  \sqrt{{\tilde \mu}^2 - m^2}$ and since the quark number density grows with $\mu$ the Fermi momentum $p_F^0$  cannot be reached, preventing the NJL results  to converge to the pQCD predictions at arbitrarily high baryonic densities.  Nevertheless, as proposed in Ref. \cite {letter}, one can assure ${\tilde \mu} \to \mu$ (and  $\sqrt{{\tilde \mu}^2 - M^2} \to  p_F^0$) by requiring $G_V(\mu) \to 0$ after the chiral transition takes place according to 
\begin{equation}
G_V(\mu) = \frac{G_V(0)} { 1 + e^{(\mu - \mu_0)/\delta}} \;,
\label{run}
\end{equation}
where $G_V(0) =G_S/3$ \cite {sugano,tulio,letter}. Considering the parametrization adopted here one has $M(0) = 367.7\,\MeV$ \cite{buballa}  and $\Lambda = 602.3\,\MeV$ so that $\mu_0 = [M(0)+\Lambda]/2\equiv 485 \, {\rm MeV}$.  The ``thickness"  $\delta = 10\, {\rm MeV}$ assures that the drop starting at $\mu=M(0)$ terminates at $\mu = \Lambda$ just as in the $N_f=2$ case \cite {letter}.  It is  obvious from Eq. (\ref{run}) that such running coupling interpolates between the two extrema, $G_V=0$ and $G_V =G_S/3$, which respectively give a softer  and a stiffer  EoS \cite{buballa, fukushima1,fukushima2}. Fig. \ref{Fig1} shows the running of $G_V(\mu)$ and also illustrates how it affects the Fermi momentum. From the physical point of view it is important to notice that the {\it ansatz}  assumes  that after chiral symmetry gets (partially) restored the repulsion among the (bare) quarks decreases  as the density increases. In other words, it is assumed that quarks with large effective masses tend to strongly repel each other as compression increases and the quark condensates decrease. After the chiral transition occurs, and the effective masses tend to their bare values, quarks can be further compressed without repelling each other indicating that repulsion should be necessary only as long as the quark condensates, $\sigma_f$, are non-zero.  A pictorial representation of the physical process driving the running of $G(\mu)$ is presented in \ref{Fig2}. Also, remark that $\delta$ was chosen   so as to give a smooth transition within a narrow $10\,{\rm MeV}$ width since taking $\delta \to 0$ could lead to discontinuities in $V_s^2$ which do not seem to be observed in the simulations of Refs.  \cite {sinansimulation,michal, weisesimulation}. Note that in order for $\sqrt{{\tilde \mu}^2 - M^2} \to \sqrt{{\mu}^2 - m^2}$ it is not compulsory that $G_S$ and $K$ run with $\mu$ since the quark condensates, multiplying these parameters in Eq.   (\ref{mass}), naturally decrease with $\mu$. In summary, for a given flavor {\it i}, $G_S$ and $K$ always appear in combinations such as $G_S \sigma_i$ and $K \sigma_j \sigma_k$ (see Eq. (\ref{mass})) which  tend to vanish at high-$\mu$ while $G_V$ appears in combinations such as $G_V n$ (see Eq. (\ref{mutilde}))  which always give a finite  high-$\mu$ contribution  when $G_V$ is fixed. 

Finally, it must be pointed out that the idea of considering $G_V$ to depend on a control parameter, as proposed here, is not new. A similar course of action was  originally taken by Kunihiro \cite {kunihiro}, who considered $G_V$ to be temperature dependent   in order to evaluate quark susceptibilities at high-$T$ (see also Refs. \cite {lorenzo,baym}).

 \section{Numerical results}
 \label{sec4}
 
\begin{figure}
\centerline{ \epsfig{file=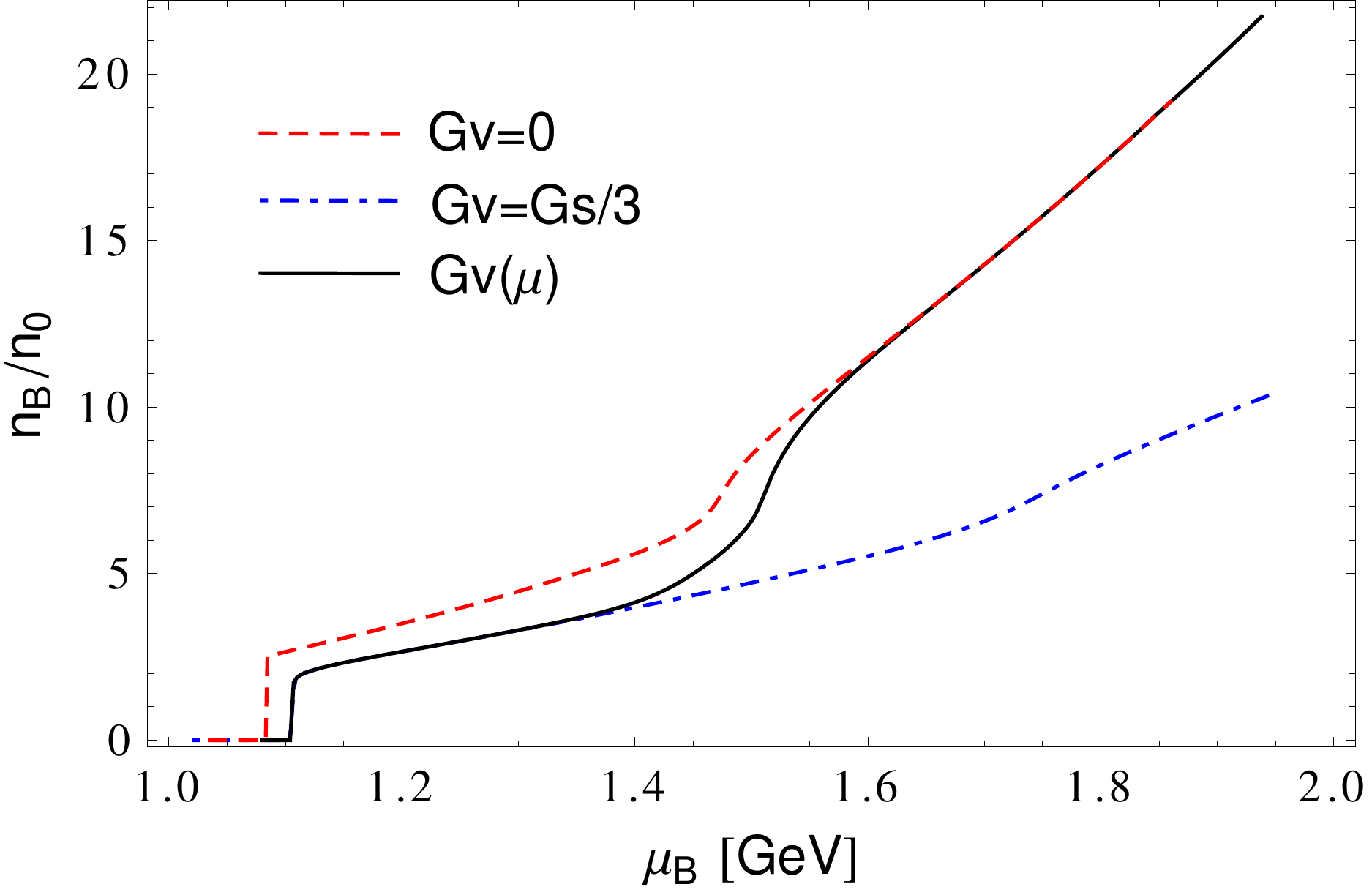,width=0.45\linewidth,angle=0}}
\caption{Baryonic number density, in units of $n_0=0.16 \, {\rm fm}^{-3}$, as a function of the baryonic chemical potential, $\mu_B = 3\mu$. The  $G_V(\mu)$ result interpolates between those predicted by $G_V=0$ and $G_V = G_S/3$. The chiral first order phase transition takes place at $\mu_B =1.083\, {\rm GeV}$ for $G_V=0$ and  at $\mu_B =1.104 \, {\rm GeV}$ for the other two cases. }
\label{Fig5}
\end{figure}

\begin{figure}
\centerline{ \epsfig{file=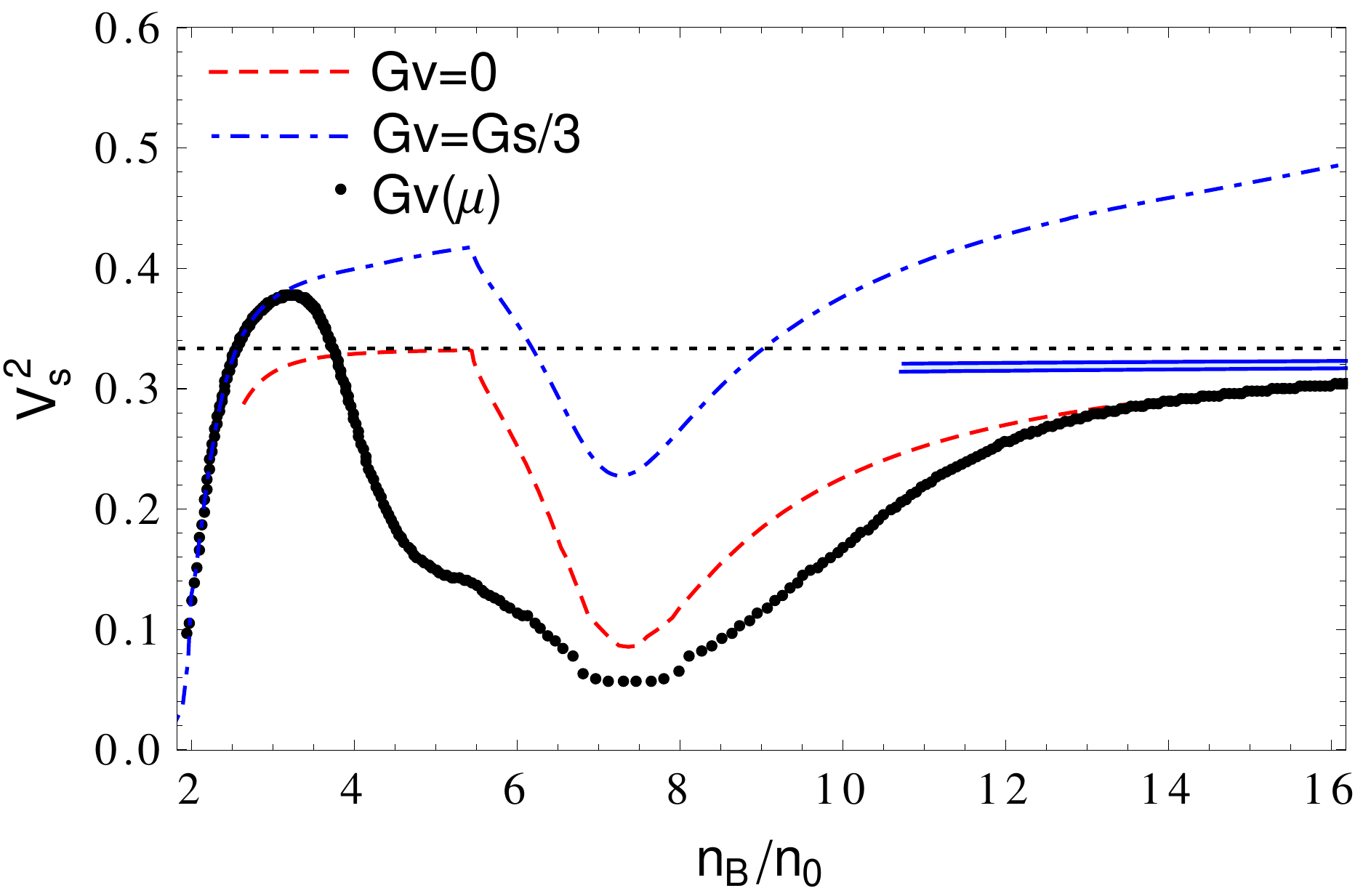,width=0.45\linewidth,angle=0}}
\caption{Speed of sound (squared) as a function of $n_B/n_0$. The running coupling predicts a non-conformal peak, $V_s^2 = 0.38$, at $n_B= 3.23 n_0$. The onset of strangeness takes place at $n_B \simeq 5.50 \, n_0$. The light  band corresponds to the pQCD results \cite {eduardo} when the $\overline {\rm MS}$ renormalization scale varies from the central value, $2 \mu$ (bottom edge), to $4\mu$ (top edge). The thin dotted line represents the conformal result, $V_s^2=1/3$. }
\label{Fig6}
\end{figure}

 Let me now analyze the effect of $G_V(\mu)$ on some relevant thermodynamical observables starting with the quark condensates for the light flavors. Fig. \ref {Fig3} shows the results for $\sigma_{u,d} = \langle {\overline u} u \rangle =\langle {\overline d} d \rangle$ obtained with different $G_V$ values. Around $\mu = 1.3 \, \mu_c$ the $G_V(\mu)$ curve, which initially agrees with the $G_V=G_S/3$ result,  converges towards the one predicted by using $G_V=0$. In the same spirit Fig. \ref {Fig4} compares the dressed Fermi momentum for light quarks, $p_F^*=\sqrt{{\tilde \mu}^2 - M^2}$, with its bare counterpart,  $p_F^0= \sqrt{\mu^2 - m^2}$, reproducing the pictorial view (shown in Fig. \ref {Fig2}) from  a quantitative perspective. Next, one can examine the baryonic number density, $n_B$, which in the present work represents the fundamental thermodynamical quantity. The result obtained with $G_V(\mu)$ is presented in Fig. \ref {Fig5} together with the predictions from the $G_V=0$ and $G_V =G_S/3$ cases. The results from $G_V(\mu)$ and $G_V=G_S/3$ agree up to $\mu_B \approx 1.4 \, {\rm GeV}$ when the former starts to agree with the $G_V=0$ curve. Fig. \ref {Fig6} shows the speed of sound squared for the three relevant cases. The $G(\mu)$ curve peaks at $n_B \simeq 3.23\, n_0 = 0.52 \, {\rm fm}^{-3}$ (corresponding to ${\cal E} = 0.59 \, {\rm GeV fm^{-3}}$) producing the non-conformal result $V_s^2 \simeq 0.38$ (note that these numerical values are consistent with those reported in Ref \cite{michal}).

The curve  then dives into the sub-conformal region reaching  $V_s^2 \simeq 0.08$ at $n_B \simeq 7.50\, n_0$ before converging to the conformal result as $n_B$ further increases.  On the other hand the dip observed in the curves of fixed $G_V$ is solely a byproduct of the onset of strangeness. This can be better understood by looking at Fig. \ref {Fig5} which displays a sudden increase of slope when $\mu$ becomes greater than $M_s(0) = 549.5 \, {\rm MeV}$. This increase in $d n_B/d\mu_B$ means that $V_s^2$ decreases (the EoS softens) as Eq. (\ref {sound}) shows. Remark that no such a   dip appears, in the case of fixed $G_V$, when the two-flavor model is considered \cite {letter}.
As a consequence of the strangeness onset the $G_V=G_S/3$ case also predicts a non-conformal peak at $n_B \simeq 5.50\, n_0$ and $V_s^2 \simeq 0.43$ while $G_V=0$ predicts a peak at $n_B \simeq 5.50\, n_0$  and $V_s^2 \simeq 0.33$. However, the value at which these peaks occur is much higher than the one predicted in Ref. \cite {michal}. Moreover, at higher $n_B$ values the use of a non-zero fixed coupling prevents convergence towards to conformal result, as the figure shows. This result is not unexpected since, as already discussed, the Fermi momentum for this case does not converge to its  pQCD counterpart, $p_F^0$. It is important to remark that the conjectured coupling running predicts that after peaking at the super-conformal region, $V_s^2$ approaches the conformal value from below, like pQCD. It should be also emphasized that the shape of the  curve generated with  $G_V(\mu)$ resembles some of those recently predicted in Refs. \cite {sinansimulation,michal,weisesimulation}. 

 \begin{figure}
\centerline{ \epsfig{file=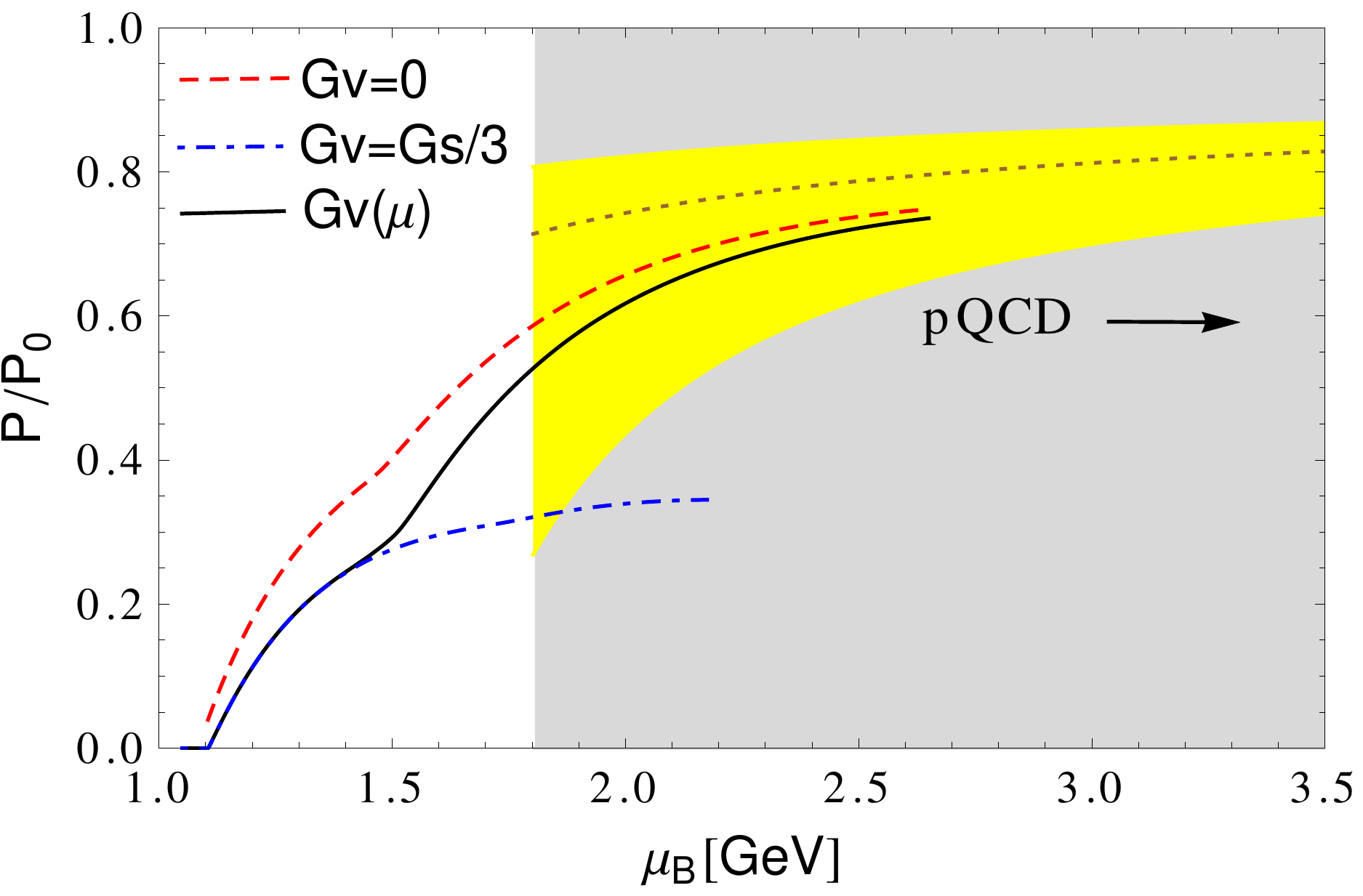,width=0.45\linewidth,angle=0}}
\caption{Pressure, normalized by  $P_0$ (see text), as a function of $\mu_B$. The gray band  represents the region where $\mu = \mu_B/3 > \Lambda$. The light  band corresponds to the pQCD results \cite {eduardo} when the $\overline {\rm MS}$ renormalization scale covers the range from  $\mu$ (bottom edge), to $4\mu$ (top edge). The dotted line represents the pQCD predictions at the central $\overline {\rm MS}$ scale, $2\mu$.}
\label{Fig7}
\end{figure}

\begin{figure}
\centerline{ \epsfig{file=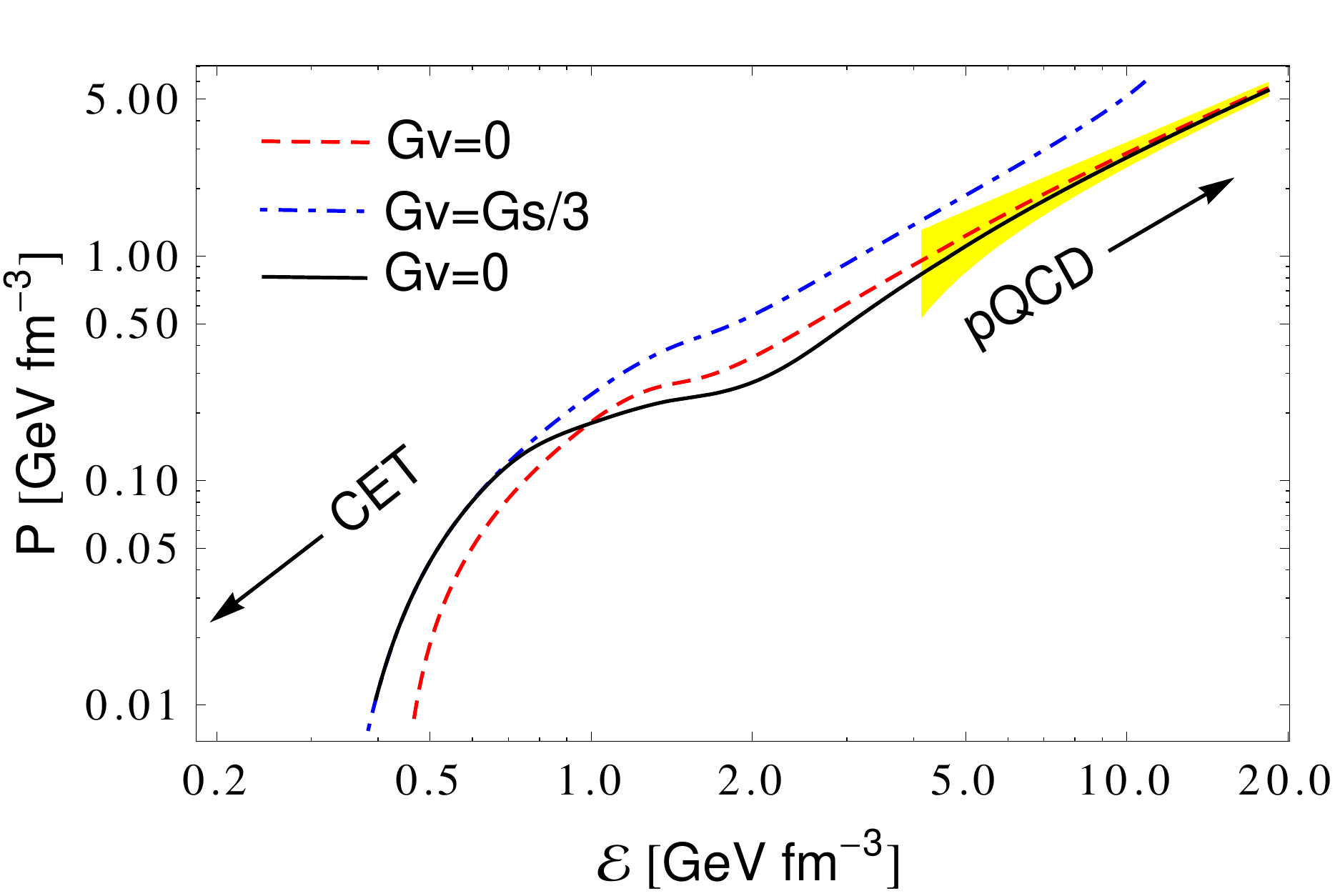,width=0.45\linewidth,angle=0}}
\caption{EoS for the three cases considered. The light  band corresponds to the pQCD results \cite {eduardo} when the $\overline {\rm MS}$ renormalization scale  varies from  $\mu$ (bottom edge) to $4\mu$ (top edge). The softening of the $G_V(\mu)$ curve takes place at ${\cal E} \simeq 0.7 \, {\rm GeV fm^{-3}}$ in agreement with Refs. \cite{nature,michal}. For completeness, the region which concerns CET has also been indicated.}
\label{Fig8}
\end{figure}

\begin{figure}
\centerline{ \epsfig{file=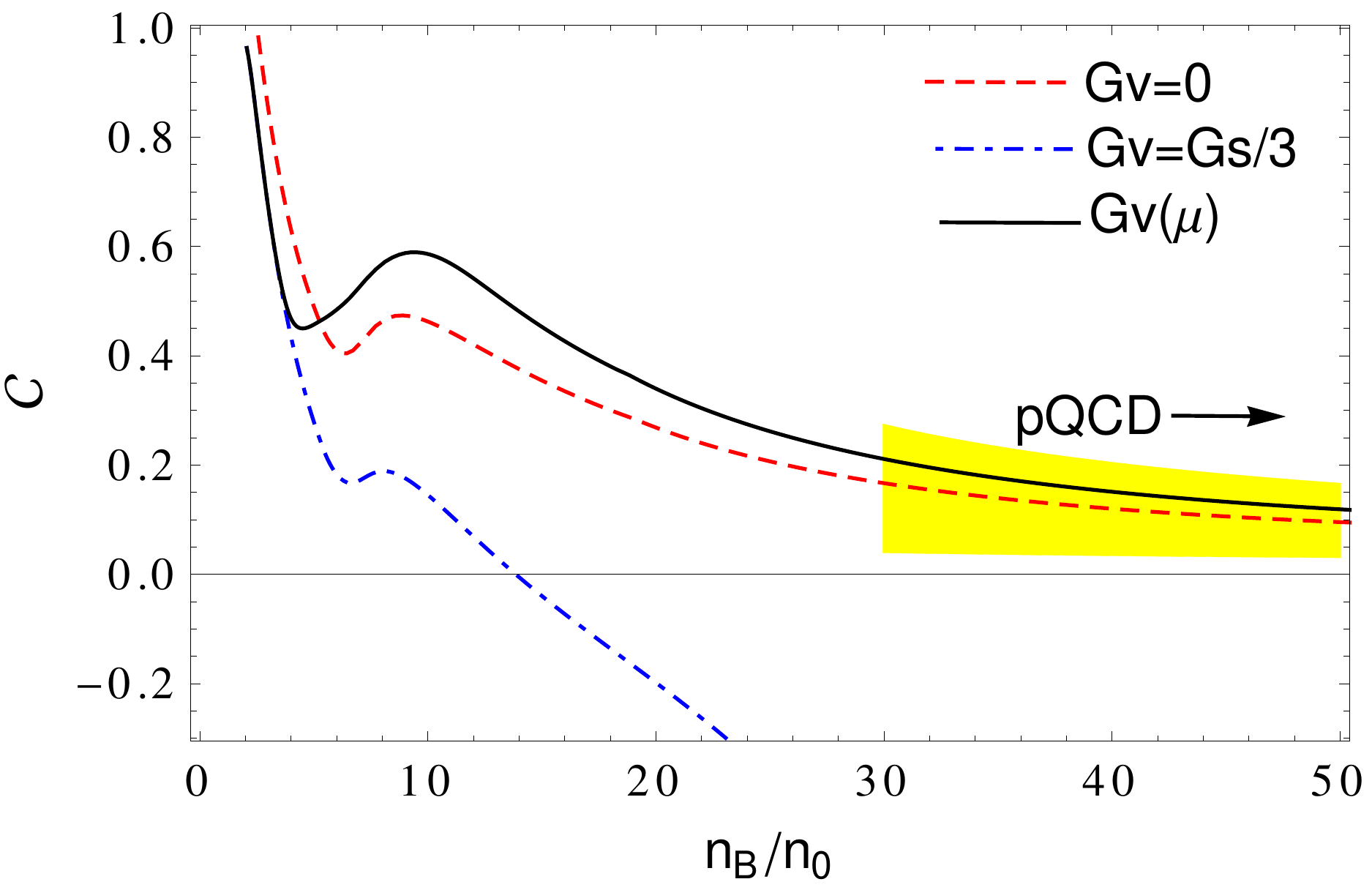,width=0.45\linewidth,angle=0}}
\caption{ Conformal measure, ${\cal C}= \Delta/{\cal E}$, as a function of $n_B/n_0$. The running coupling predicts a  change of slope of high amplitude, at $n_B = 4-10 \, n_0$. A fixed $G_V$ leads to a negative $\cal C$ at $n_B/n_0 \simeq 14$. The light  band corresponds to the pQCD results \cite {eduardo} when the $\overline {\rm MS}$ renormalization scale covers the range from  $\mu$ (top edge) to $4\mu$ (bottom edge).}
\label{Fig9}
\end{figure}

The NJL pressure together with the pQCD results for the $N_f=2+1$ case, obtained from Ref. \cite {eduardo}, is displayed in Fig. \ref{Fig7}. The pQCD results were obtained by varying the  $\overline {\rm MS}$ renormalization scale from  $\mu$  to $4\mu$ while the Fermi-Dirac limit for free massless quarks, used to normalize the pressure in Fig. \ref{Fig7}, reads 
 \begin{equation}
 P_0 =  \frac {N_c N_f}{12 \pi^2} \left ( \frac{\mu_B}{3} \right )^4 \,.
\end{equation} 
Fig. \ref{Fig8} displays the NJL EoS as well as the pQCD result (generated from Ref. \cite{eduardo}). The figure clearly shows that the predictions coming from $G_V(\mu)$ and $G_V=0$ agree with pQCD at high energies while $G_V=G_S/3$ does not. Of utmost importance is the fact that at ${\cal E} \simeq 0.7 \, {\rm GeV fm^{-3}}$ a sudden change of slope takes place producing the softening of the EoS produced by $G_V(\mu)$, in accordance with Refs. \cite{nature,michal}. A second change of slope happens at ${\cal E} \simeq 2 \, {\rm GeV fm^{-3}}$ redressing the  $G_V(\mu)$ curve so that it smoothly joins  the pQCD band.

Finally, let me use the proposed  model in order to examine the possibility that the  conformal measure, $\cal C$, remains positive at all densities. This important question has been recently addressed in Ref.  \cite{fukushimatrace} where the authors  have considered 
the trace anomaly, $\Delta$, which  trivially relates to $\cal C$ via ${\cal C}=\Delta/{\cal E}$.
 Fig. \ref{Fig9} shows that the fixed $G_V= G_S/3$ produces a maximally stiff EoS which yields a negative $\cal C$ for $n_B \gtrsim 14\, n_0$. When repulsion is absent, the EoS is softer causing ${\cal C} \to 0$ as $n_B \to \infty$ in conformity with the pQCD predictions (generated from Ref. \cite {eduardo}). At the same time, our running coupling predicts a  change of slope of high amplitude, at $n_B = 4-10 \, n_0$, preventing  $\cal C$ from becoming negative. In summary, $G(\mu)$ shifts the  high-$n_B$ behavior of the trace anomaly which then approaches zero while remaining  positive, supporting the hypothesis advanced in Ref. \cite{fukushimatrace}.

 \section{Conclusions}
 \label{sec5}
 
 The three-flavor NJL model with a repulsive vector channel, parametrized by $G_V$, has been considered in  the evaluation of the EoS describing symmetric cold quark matter.  The work extends the application performed in Ref. \cite{letter}, where the two-flavor version has been considered  in the presence of density dependent repulsive coupling, $G_V(\mu)$. Here, I have  shown that the presence of strangeness does not affect the main physical properties displayed by key thermodynamical quantities evaluated with $G_V(\mu)$. The advantage of such a model is that one is then able to interpolate between a regime where repulsion is high (the EoS is stiff) and a regime where repulsion is low (the EoS is soft).  In this way the NJL model can simultaneously  observe astrophysical constraints, which require the EoS to be stiff at lower densities, while producing results which agree with pQCD at arbitrarily high densities. For instance, considering the moderate value $G_V(0)=G_S/3$ this work shows that it is possible to describe  a non-conformal peak at $V_s^2 = 0.38$ and $n_B = 3.23 \, n_0 =  0.52 \, {\rm fm}^{-3}$ (corresponding to ${\cal E} = 0.59 \, {\rm GeV fm^{-3}}$). These numerical values are in good agreement with some of the values quoted  in Ref. \cite{michal}. I have also shown that, as the density increases, the interpolating model predicts that $V_s^2$ approaches the pQCD (conformal) prediction, $V_s^2 \to 1/3$ from below, as expected. Another important result obtained here shows that the proposed model can produce a noticeable change of slope in an initially hard EoS  so that it will soften and join the pQCD predictions at higher energy densities. Interestingly enough this change happens at ${\cal E} \simeq 0.7 \,{\rm GeV fm^{-3}}$, in conformity  with predictions made in Refs. \cite {nature,michal, new}. 
The results also indicate   that a non-conformal peak in $V_s^2$ is not in tension with the trace anomaly being positive for all densities, a result which agrees with a scenario proposed in  Ref. \cite{fukushimatrace}.  As explicitly shown here, these findings   cannot be reproduced if one naively uses $G_V=0$ (the EoS is far too soft at low-$n_B$), or if one fixes $G_V$ to a finite value (the EoS is far too hard at high-$\mu_B$). 
At first sight it seems remarkable that with  a simple modification  the NJL model is able to reproduce such highly non-trivial results,  which were originally  obtained through the use of more sophisticated approaches \cite {nature,michal,fukushimatrace}. However, it should be clear that the simple modification encoded within the $G_V(\mu)$ running   has physical consequences   which in turn imply that the fundamental concept of repulsion should be altogether reviewed. More precisely, the results obtained here suggest that quarks with bare masses do not tend to repel each other when compressed, in opposition to the behavior displayed by quarks with effective masses. Obviously,  the simple {\it ansatz} proposed in the present work  is not unique  so that one is free to consider alternative forms (such as gaussian, skewed gaussian, etc) as well as the use of other parametrizations {\it provided that $G_V$  decreases} with the density after the chiral transition takes place (keeping in mind that this is the main ingredient driving the crucial change of slope observed in the corresponding EoS). In principle, the mechanism described here can be generalized to any  model which contains a repulsive channel. Possible  extensions  include the consideration of non-symmetric quark matter in $\beta$-equilibrium, in order to describe quark stars, as well as the inclusion of a diquark interaction channel, in order to explore the high-density region of QCD, among others.  In future applications one could also consider replacing  the popular pQCD predictions with those furnished by  the {\it renormalization group optimized perturbation theory}, since this resummation technique generates results which are less sensitive to scale changes \cite {rgopt1,rgopt2,rgopt3}.

\acknowledgments
 
 The author is partially supported by Conselho
Nacional de Desenvolvimento Cient\'{\i}fico e Tecnol\'{o}gico (CNPq),
Grant No  307261/2021-2  and by CAPES - Finance  Code  001.  
 This work has also been financed  in  part  by
Instituto  Nacional  de  Ci\^encia  e Tecnologia de F\'{\i}sica
Nuclear e Aplica\c c\~{o}es  (INCT-FNA), Process No.  464898/2014-5.


\begin{thebibliography}{99} 

\bibitem{njl}  Y. Nambu and G. Jona-Lasinio, Phys. Rev. \textbf{122}, 345
  (1961); {\it ibid.} \textbf{124}, 246 (1961).

\bibitem{mit1} A. Chodos, R.L. Jaffe, K. Johnson, C.B. Thorn and V.F. Weisskopf, Phys. Rev. D {\bf 9}, 3471 (1974).

\bibitem{mit2} A. Chodos, R.L. Jaffe, K. Johnson and C.B. Thorn, Phys. Rev. D {\bf 10}, 2599 (1974).

\bibitem{mit3} T. DeGrand, R.L. Jaffe, K. Johnson and J. Kiskis, Phys. Rev. D {\bf 12}, 2060 (1975).

\bibitem{buballa} M.~Buballa, Phys. Rept. \textbf{407}, 205 (2005), [\href{https://arxiv.org/abs/hep-ph/0402234 }{{\tt
  arXiv:hep-ph/0402234 }}].
  
\bibitem{cet1} S.~Gandolfi, A.~Y.~Illarionov, K.~E.~Schmidt, F.~Pederiva and S.~Fantoni,
%``Quantum Monte Carlo calculation of the equation of state of neutron matter,''
Phys. Rev. C \textbf{79}, 054005 (2009),
[\href{https://arxiv.org/abs/0903.2610}{{\tt
  arXiv:0903.2610}}].
%doi:10.1103/PhysRevC.79.054005


\bibitem{cet2} I.~Tews, T.~Kr\"uger, K.~Hebeler and A.~Schwenk,
%``Neutron matter at next-to-next-to-next-to-leading order in chiral effective field theory,''
Phys. Rev. Lett. \textbf{110},  032504 (2013),
[\href{https://arxiv.org/abs/1206.0025}{{\tt
  arXiv:1206.0025}}].

%doi:10.1103/PhysRevLett.110.032504
%[arXiv:1206.0025 [nucl-th]].


\bibitem{pqcd1} B. A. Freedman and L. D. McLerran, Phys. Rev. D \textbf{16}, 1169 (1977).

\bibitem{pqcd2}  A.~Kurkela, P.~Romatschke and A.~Vuorinen,
%``Cold Quark Matter,''
Phys. Rev. D \textbf{81}, 105021 (2010),
[\href{https://arxiv.org/abs/0912.1856}{{\tt
  arXiv:0912.1856}}].
%doi:10.1103/PhysRevD.81.105021
%[arXiv:0912.1856 [hep-ph]].

\bibitem{pqcd3} T.~Gorda, A.~Kurkela, P.~Romatschke, M.~S\"appi and A.~Vuorinen,
%``Next-to-Next-to-Next-to-Leading Order Pressure of Cold Quark Matter: Leading Logarithm,''
Phys. Rev. Lett. \textbf{121}, 202701 (2018),
[\href{https://arxiv.org/abs/1807.04120}{{\tt
  arXiv:1807.04120}}].
%doi:10.1103/PhysRevLett.121.202701
%1807.04120[arXiv:1807.04120 [hep-ph]].


\bibitem{kojo} T.~Kojo, APPS Bull. \textbf{31}, 11 (2021), 
[\href{https://arxiv.org/abs/2011.10940}{{\tt
  arXiv:2011.10940}}].



\bibitem{nature} E.~Annala, T.~Gorda, A.~Kurkela, J.~N\"attil\"a and A.~Vuorinen,
%``Evidence for quark-matter cores in massive neutron stars,''
Nature Phys. \textbf{16},  907 (2020),
[\href{https://arxiv.org/abs/1903.09121}{{\tt
  arXiv:1903.09121}}].

%doi:10.1038/s41567-020-0914-9
%[arXiv:1903.09121 [astro-ph.HE]].


\bibitem{astro1} J. Antoniadis et al., Science \textbf{ 340}, 6131 (2013), 
[\href{https://arxiv.org/abs/1304.6875}{{\tt
  arXiv:1304.6875}}].

\bibitem{astro2} H. T. Cromartie et al. (NANOGrav), Nature Astron. \textbf{4}, 72
(2019), [\href{https://arxiv.org/abs/1904.06759}{{\tt
  arXiv:1904.06759}}].



\bibitem{astro3} E. Fonseca et al., Astrophys. J. Lett. \textbf{915}, L12 (2021),
[\href{https://arxiv.org/abs/2104.00880}{{\tt
  arXiv:2104.00880}}].
  
 \bibitem{measure1} B. Margalit and B. D. Metzger, Astrophys. J. Lett. \textbf{850}, L19
(2017),  [\href{https://arxiv.org/abs/1710.05938}{{\tt
  arXiv:1710.05938}}].

\bibitem{measure2} L. Rezzolla, E. R. Most, and L. R. Weih, Astrophys. J. Lett.
\textbf{852}, L25 (2018), 
[\href{https://arxiv.org/abs/1711.00314}{{\tt
  arXiv:1711.00314}}].

\bibitem{measure3} M. Ruiz, S. L. Shapiro, and A. Tsokaros, Phys. Rev. D \textbf{97},
021501 (2018), [\href{https://arxiv.org/abs/1711.00473}{{\tt
  arXiv:1711.00473}}].

\bibitem{measure4} M. Shibata, E. Zhou, K. Kiuchi, and S. Fujibayashi, Phys. Rev.
D \textbf{100}, 023015 (2019), 
[\href{https://arxiv.org/abs/1905.03656}{{\tt
  arXiv:1905.03656}}].

\bibitem{measure5} A.~Nathanail, E.~R.~Most and L.~Rezzolla,
Astrophys. J. Lett. \textbf{908}, L28 (2021),  [\href{https://arxiv.org/abs/2101.01735}{{\tt
  arXiv:2101.01735}}].

\bibitem{sinansimulation}S.~Altiparmak, C.~Ecker and L.~Rezzolla, 
[\href{https://arxiv.org/abs/2203.14974}{{\tt
  arXiv:2203.14974}}].
  
\bibitem{michal}   M.~Marczenko, L.~McLerran, K.~Redlich and C.~Sasaki, [\href{https://arxiv.org/abs/2207.13059}{{\tt
  arXiv:2207.13059}}].  
  
\bibitem{weisesimulation}
L.~Brandes, W.~Weise and N.~Kaiser, [\href{https://arxiv.org/abs/2208.03026}{{\tt
  arXiv:2208.03026}}].

  
  
\bibitem{fukushima1} K.~Fukushima, Phys. Rev. D \textbf{77}, 114028 (2008),
[erratum: Phys. Rev. D \textbf{78}, 039902 (2008)], [\href{https://arxiv.org/abs/0803.3318  }{{\tt
  arXiv:0803.3318  }}].

\bibitem{fukushima2} K.~Fukushima, Phys. Rev. D \textbf{78}, 114019 (2008),
[\href{https://arxiv.org/abs/0809.3080}{{\tt
  arXiv:0809.3080}}].  
  

 \bibitem{letter} M.~B.~Pinto,
%``Repulsive vector interaction as a trigger for the non-conformal peak in $V_s^2$,''
[\href{https://arxiv.org/abs/2208.06911}{{\tt
  arXiv:2208.06911}}]. 
  
\bibitem{fukushimatrace}
Y.~Fujimoto, K.~Fukushima, L.~D.~McLerran and M.~Praszalowicz, [\href{https://arxiv.org/abs/2207.06753}{{\tt
  arXiv:2207.06753}}].

\bibitem{campoB} D.~P.~Menezes, M.~B.~Pinto, L.~B.~Castro, P.~Costa and C.~Provid\^encia,
%``Repulsive Vector Interaction in Three Flavor Magnetized Quark and Stellar Matter,''
Phys. Rev. C \textbf{89}, 055207 (2014), [\href{https://arxiv.org/abs/1403.2502}{{\tt
  arXiv:1403.2502}}]. 
%doi:10.1103/PhysRevC.89.055207
%[arXiv:1403.2502 [nucl-th]].  
  
\bibitem{rkh} P.~Rehberg, S.~P.~Klevansky and J.~H\"{u}fner,
%``Hadronization in the SU(3) Nambu-Jona-Lasinio model,''
Phys. Rev. C \textbf{53}, 410 (1996), 
[\href{https://arxiv.org/abs/9506436}{{\tt
  arXiv:9506436}}].
%doi:10.1103/PhysRevC.53.410
%[arXiv:hep-ph/9506436 [hep-ph]].  

\bibitem{tulio} T.~E.~Restrepo, J.~C.~Macias, M.~B.~Pinto and G.~N.~Ferrari,
Phys. Rev. D \textbf{91}, 065017 (2015), [\href{https://arxiv.org/abs/1412.3074}{{\tt
  arXiv:1412.3074}}].

\bibitem{sugano} J.~Sugano, J.~Takahashi, M.~Ishii, H.~Kouno and M.~Yahiro, Phys. Rev. D \textbf{90},  037901 (2014), [\href{https://arxiv.org/abs/1405.0103}{{\tt
  arXiv:1405.0103}}].
  
\bibitem{constanca} R.~C\^amara Pereira, P.~Costa and C.~Provid\^encia,
%``Two-solar-mass hybrid stars: a two model description with the Nambu-Jona-Lasinio quark model,''
Phys. Rev. D \textbf{94},  094001 (2016),  [\href{https://arxiv.org/abs/1610.06435}{{\tt
  arXiv:1610.06435}}].
%doi:10.1103/PhysRevD.94.094001
%[arXiv:1610.06435 [nucl-th]].
  
\bibitem{zong} H.-S. Zong and W.-M. Sun, Int. J. Mod. Phys. A \textbf{23},
3591 (2008).  

\bibitem{newtulio} T.~E.~Restrepo, C.~Provid\^{e}ncia and M.~B.~Pinto, [\href{https://arxiv.org/abs/2212.11184}{{\tt
  arXiv:2212.11184}}].

%``Non-strange quark stars within resummed QCD,''
%[arXiv:2212.11184 [hep-ph]].
  
\bibitem{kunihiro} T.~Kunihiro, Phys. Lett. B \textbf{271}, 395 (1991).

\bibitem{lorenzo}   L.~Ferroni and V.~Koch, Phys. Rev. C \textbf{83}, 045205 (2011), [\href{https://arxiv.org/abs/1003.4428}{{\tt
  arXiv:1003.4428}}].  

\bibitem{baym} Y.~Song, G.~Baym, T.~Hatsuda and T.~Kojo,
%``Effective repulsion in dense quark matter from nonperturbative gluon exchange,''
Phys. Rev. D \textbf{100}, 034018 (2019), [\href{https://arxiv.org/abs/1905.01005}{{\tt
  arXiv:1905.01005}}].  

%doi:10.1103/PhysRevD.100.034018
%[arXiv:1905.01005 [astro-ph.HE]].  
  
\bibitem{eduardo} E.~S.~Fraga and P.~Romatschke, Phys. Rev. D \textbf{71}, 105014 (2005), [\href{https://arxiv.org/abs/hep-ph/0412298}{{\tt
  arXiv:hep-ph/0412298}}].
  
\bibitem{new} O.~Komoltsev, [\href{https://arxiv.org/abs/2208.03086}{{\tt
  arXiv:2208.03086}}].  
  
\bibitem{rgopt1} J.~L.~Kneur, M.~B.~Pinto and T.~E.~Restrepo, Phys. Rev. D \textbf{100}, 114006 (2019), [\href{https://arxiv.org/abs/1908.08363}{{\tt
  arXiv:1908.08363}}].
  
\bibitem{rgopt2} J.~L.~Kneur, M.~B.~Pinto and T.~E.~Restrepo,
%``QCD pressure: Renormalization group optimized perturbation theory confronts lattice,''
Phys. Rev. D \textbf{104},  L031502 (2021), 
[\href{https://arxiv.org/abs/2101.02124}{{\tt
  arXiv:2101.02124}}].

%doi:10.1103/PhysRevD.104.L031502
%[arXiv:2101.02124 [hep-ph]].  

\bibitem{rgopt3} J.~L.~Kneur, M.~B.~Pinto and T.~E.~Restrepo,
%``Renormalization group improved pressure for hot and dense quark matter,''
Phys. Rev. D \textbf{104},  034003 (2021),
[\href{https://arxiv.org/abs/2101.08240}{{\tt
  arXiv:2101.08240}}].
%doi:10.1103/PhysRevD.104.034003
%[arXiv:2101.08240 [hep-ph]].


\end{thebibliography}
\end{document}